\documentclass[preprint,12pt,longnamesfirst]{aastex}

\usepackage{latexsym}
\shorttitle{X-rays from WR in the MCs}
\shortauthors{Guerrero \& Chu}

\newcommand{\kms}{\mbox{km~s$^{-1}$}}

\begin{document}

\title{An X-ray Survey of Wolf-Rayet Stars in the Magellanic Clouds. \\
       I. The Chandra ACIS Dataset}

\author{Mart\'{\i}n A.\ Guerrero\altaffilmark{1,2}, 
        You-Hua Chu\altaffilmark{2}} 
\altaffiltext{1}{Instituto de Astrof\'{\i}sica de Andaluc\'{\i}a, CSIC, 
        Apdo.\ 3004, 18080, Granada, Spain}
\altaffiltext{2}{Astronomy Department, University of Illinois, 
        1002 W. Green Street, Urbana, IL 61801, USA}

\begin{abstract}

Wolf-Rayet (WR) stars are evolved massive stars with strong fast stellar 
winds.  
WR stars in our Galaxy have shown three possible sources of X-ray 
emission associated with their winds: shocks in the winds, colliding 
stellar winds, and wind-blown bubbles; however, quantitative analyses 
of observations are often hampered by uncertainties in distances and 
heavy foreground absorption.  
These problems are mitigated in the Magellanic Clouds (MCs), which are
at known distances and have small foreground and internal extinction.
We have therefore started a survey of X-ray emission associated with
WR stars in the MCs using archival \emph{Chandra}, \emph{ROSAT}, and
\emph{XMM-Newton} observations.  
In the first paper of this series, we report the results for 70 WR 
stars in the MCs using 192 archival \emph{Chandra} ACIS observations.
X-ray emission is detected from 29 WR stars.
We have investigated their X-ray spectral properties, luminosities, and 
temporal variability.  
These X-ray sources all have luminosities greater than a few times 
10$^{32}$ ergs~s$^{-1}$, with spectra indicative of highly absorbed 
emission from a thin plasma at high temperatures typical of colliding 
winds in WR+OB binary systems.  
Significant X-ray variability with periods ranging from a few 
hours up to $\sim$20 days is seen associated with several WR 
stars.  
In most of these cases, the X-ray variability can be linked to the 
orbital motion of the WR star in a binary system, further supporting 
the colliding wind scenario for the origin of the X-ray emission 
from these stars.

\end{abstract}


\section{Introduction}

Hot, massive stars are so luminous that their radiation can drive 
fast stellar winds with terminal velocities ($v_\infty$) of 
1,000--3,000 \kms\ \citep{PBH90}.  
Fast stellar winds have been detected from main-sequence O stars, 
as well as evolved blue supergiants.
The most powerful fast winds belong to Wolf-Rayet (WR) stars, which 
are evolved massive stars with their H-rich envelopes stripped off.
WR stars have typical mass loss rates ($\dot{M}$) of a few 
$\times10^{-5}$ $M_\odot$ yr$^{-1}$ \citep{dJNvH88} and stellar 
wind mechanical luminosities ($L_{\rm w}$ $\equiv$ $\frac{1}{2} 
\dot{M} v_\infty^2$) of $10^{37}-10^{38}$ ergs s$^{-1}$.
The powerful stellar winds of WR stars are associated with three
types of shocks that can produce X-ray emission:
shocks in the wind itself, colliding winds in a binary system, 
and shocked wind in a circumstellar bubble.

Shocks in the wind itself are produced by stochastic or 
radiatively induced instabilities and the post-shock gas 
reaches X-ray-emitting temperatures \citep{LW80,GO95}.
Such X-ray emission has been detected, for example, from the WN4 
WR star HD\,50896 \citep{WS96}.  
This is basically the same X-ray emission mechanism for O and 
early B stars \citep{Betal97}, but WR winds are heavily enriched 
in metals (C, N, O) so that their X-ray emission can be highly 
absorbed \citep{P87}.  
The X-ray emission from shocks in a wind appears as an 
unresolved point source, and its spectral shape can be 
described by thermal plasma emission at temperatures 
of a few $\times$10$^6$~K \citep{Setal02b}.

In a WR+OB binary system, the WR wind collides with the companion's 
fast wind and generates shock-heated plasma at the interaction region.  
The X-ray emission from such an interaction region has been resolved
in the binary system of WR\,147 \citep{Petal02}.  
Observations of V444 Cygni \citep{Cetal96} illustrate that the
physical conditions and X-ray luminosity of the hot gas at the 
collision zone vary with the orbital phase of the binary system.
X-ray emission from colliding winds typically shows plasma 
temperatures $> 10^7$~K \citep{C03}.

The fast WR wind can blow a bubble in the ambient medium, and 
the bubble interior is filled with shocked stellar wind
that emits X-rays \citep{GML96,GLM96}.  
X-ray emission from a WR bubble is distributed and is expected to
peak near the inner wall of the bubble shell.
Diffuse X-ray emission of hot interior gas has been detected
from only two WR bubbles, NGC\,6888 and S\,308 
\citep{B88,WWW94,W99,Cetal03}.
Their X-ray spectra indicate plasma temperatures of 
(1--2) $\times 10^6$~K \citep{CGG03}.

Systematic X-ray surveys of WR stars using \emph{Einstein} and \emph{ROSAT} 
observations have been limited to our Galaxy \citep{P87,PHC95,W96}.  
The study of the X-ray properties of Galactic WR stars is difficult 
because the high extinction in the Galactic plane hampers the detection 
itself, uncertain distances result in poorly determined luminosities, 
and the unknown existence of binary companions confuses the assessment 
of origin of X-ray emission.  
The limited spatial resolution and sensitivity of \emph{Einstein} and 
\emph{ROSAT} have produced a number of spurious detections, as 
pointed out by \citet{Oetal03}.

The \emph{Chandra} and \emph{XMM-Newton} X-ray observatories, with their 
unprecedented angular resolution and sensitivity, make it possible to 
study the X-ray emission from WR stars in the nearby Magellanic Clouds 
(MCs), as illustrated by the recent works on the 30 Doradus region in 
the Large Magellanic Cloud (LMC) by \citet{PPL02} and \citet{Tetal06}, 
and on the N\,66 region in the Small Magellanic Cloud (SMC) by 
\citet{Netal02}.  
The MCs have typical foreground reddening of $E_{B-V}$$\sim$0.04--0.09 
\citep{SI91} and internal reddening of $E_{B-V}$$\sim$0.06 
\citep{B91}, much smaller than the reddening in the Galactic 
plane, thus making it easier to detect the soft X-ray 
emission from single WR stars and WR bubbles.  
Furthermore, WR stars in the MCs are at known distances, 50 kpc for 
the LMC \citep{F99} and 60 kpc for the SMC \citep{Cetal00}, so their 
X-ray luminosities can be determined without the uncertainties for 
Galactic WR stars.  
More importantly, there have been systematic spectroscopic surveys 
for all WR stars in the MCs to search for binaries \citep{BMN01, 
FMG03a, FMG03b, Setal03}, providing an invaluable database to aid 
the interpretation of X-ray emission from WR stars.

We have started a search for X-ray emission from WR stars in the MCs 
using \emph{Chandra}, \emph{XMM-Newton}, and \emph{ROSAT} archival 
observations.   
In this first paper of a series, we report our \emph{Chandra} 
ACIS archival search for X-ray emission from WR stars in the MCs.  
At the distance of the MCs, the superb angular resolution of 
\emph{Chandra} is especially suited to resolve point X-ray sources 
from surrounding diffuse emission or nearby point sources, allowing 
us to make credible associations of X-ray sources with WR stars.  
Subsequent papers of this series will report our underway analysis 
of \emph{XMM-Newton} and \emph{ROSAT} archival observations of WR 
stars in the MCs.

\section{Chandra ACIS Observations of WR stars in the MCs}

The \emph{Chandra} Archive\footnote{
The \emph{Chandra} Archive is available using the \emph{Chandra} Search 
and Retrieval Interface (\emph{ChaSeR}) at the \emph{Chandra} X-ray 
Observatory site (http://cxc.harvard.edu).
}
available by 2004 October was used to search for observations that 
contained WR stars included in the compilations by \citet{BAT99} for 
the LMC and by \citet{MOP03} for the SMC.
Only observations made with the Advanced CCD Imaging Spectrometer 
(ACIS) were included in this search because of its sensitivity 
and spectral resolution.
We first selected ACIS observations whose nominal pointings were
within 30$\arcmin$ from a WR star, then retrieved the data from
the \emph{Chandra} Archive to check whether the WR star was
actually included within the ACIS's non-circular field of view,
and to identify the CCD chip that registered the WR star.
This search resulted in 192 useful \emph{Chandra} ACIS 
observations of 61 WR stars in the LMC and 9 WR stars 
in the SMC.  
These observations are listed in Tables~1 and 2 for the LMC
and SMC, respectively. 
The table columns are: 
(1) the WR catalog number \citep{BAT99,MOP03},
(2) WR star's common name, 
(3) and (4) the equatorial coordinates for J2000.0 as given by \citet{BAT99} 
for the LMC WR stars and by \citet{MOP03} for the SMC WR stars, 
(5) \emph{Chandra} observation ID, 
(6) instrument at \emph{Chandra}'s aimpoint, 
(7) identification of the ACIS CCD that registered the WR star, 
(8) exposure time of the observation, $t_{\rm exp}^{\rm obs}$, and 
(9) usable exposure time, $t_{\rm exp}$, as explained below.

The data reduction and analysis of these observations were 
performed using the \emph{Chandra} X-ray Center software 
CIAO v3.1 and the HEASARC FTOOLS and XSPEC v11.0.1 routines 
\citep{A96}.  
In order to build a homogeneous database and to correct the charge 
transfer inefficiency (CTI) effects, the level 1 event files of all 
observations were reprocessed using CIAO tasks to apply the most 
up-to-date calibration files available in the calibration database 
CALDB v2.28.  
The CTI correction is important for datasets obtained with 
front-illuminated (FI) CCDs, while datasets obtained with the 
back-illuminated (BI) CCDs, i.e., ACIS-S1 and S3, show less 
noticeable CTI effects that have not been corrected.
Using CIAO v3.1 and CALDB 2.28, the CTI correction can only be 
applied to FI datasets taken at a focal plane temperature of 
$-$120$^\circ$C.  
For FI datasets obtained at temperatures of $-$110$^\circ$C, 
we used instead the CTI corrector v1.38 \citep{Tetal00}, while 
no correction has been applied to FI datasets at temperatures 
of $-$100$^\circ$C\footnote{
A limited number of FI datasets at temperatures of $-$100$^\circ$C 
are used in the spectral analysis of WR stars that are detected in 
X-rays.  
Among these WR stars, only 5 had observations obtained with FI chips 
at temperatures of $-$100$^\circ$C, with contributions to the total 
exposure time $<$5\%, but in the case of LMC-WR\,82, for which the 
exposure time of the FI datasets at $-$100$^\circ$C is $\sim$12\% of 
the total exposure time.  
}. 
All reprocessed level 1 event lists were subsequently filtered to 
select events with good \emph{ASCA} grades and clean status, and 
to reject bad aspect intervals.  
Known aspect offsets were corrected to improve the absolute astrometry 
of the data to be better than 1\arcsec.   
Finally, we removed the time intervals when the background count 
rate was 20\% above the quiescent mean value \citep{M01}.  
The remaining intervals of good observation yielded the usable
exposure time, $t_{\rm exp}$, listed in Column 9 of Tab.~1 and 
2.  
Several WR stars have multiple short \emph{Chandra} observations.
For these stars, the reprocessed level 2 event files obtained with 
the same instrument were merged to improve the signal-to-noise 
ratio, increasing the sensitivity of the search for X-ray emission 
from these stars.

\section{Results}

\subsection{Search for X-ray Emission from WR Stars in the MCs}

To search for X-ray emission from WR stars using the \emph{Chandra} 
ACIS observations listed in Tables 1 and 2, we first examined the 
X-ray and optical images of the WR stars.  
The X-ray images extracted from the \emph{Chandra} ACIS observations 
include all events in the 0.3-7.0 keV energy band and have been 
constructed with a pixel size from 0\farcs5 to 1\farcs5, depending 
on the signal-to-noise ratio and the density of sources in the field 
of view.  
These images are further smoothed using a Gaussian profile with 
a FWHM of 1--5 pixels.  
Optical images extracted from the Digitized Sky Survey\footnote{
The Digitized Sky Survey (DSS) is based on photographic data obtained 
using the UK Schmidt Telescope and the Oschin Schmidt Telescope on 
Palomar Mountain.  
The UK Schmidt was operate by the Royal Observatory of Edinburgh, with 
funding from the UK Science and Engineering Research Council, until 1988 
June, and thereafter by the Anglo-Australian Observatory. 
The Palomar Observatory Sky Survey was funded by the National Geographic 
Society. 
The Oschin Schmidt Telescope is operated by the California Institute of 
Technology and Palomar Observatory. 
The plates were processed into the present compressed digital form with 
the permission of these institutes. 
The Digitized Sky Survey was produced at the Space Telescope Science 
Institute under US government grant NAGW-2166.} (DSS) have been used to 
identify the WR stars in the optical images using the coordinates 
provided by \citet{BAT99} for the LMC and by \citet{MOP03} for the SMC.  
Finally, the X-ray and optical images have been compared to search for 
a point-like X-ray source at the location of a WR star or for diffuse 
emission from a WR star's circumstellar bubble.  
This procedure has provided us with a number of point-like X-ray 
sources associated with WR stars, but no localized diffuse X-ray 
emission indicative of a wind-blown bubble is detected around any 
WR star in the MCs.

In order to perform a quantitative analysis, we define source and 
background apertures for each WR star, compute the background-subtracted 
counts within the source aperture, and compare the value to that of the
local rms background.
Circular source apertures are used and their radii have been selected to 
encircle at least 90\% of the X-ray flux at $\sim$1.5 keV.  
Therefore, the radius of a source aperture increases with the off-axis 
angle of the WR star.  
Tailored, smaller source apertures are used when nearby point sources
or diffuse emission contaminates the 90\%-power aperture unevenly (e.g., 
LMC-WR\,103 and SMC-WR\,5).  
The CIAO task {\it dmextract} has been used to perform the statistics 
of the source and background apertures of each WR star, allowing us 
to derive its background-subtracted counts and local rms background.  
For WR stars with multiple observations at very different off-axis angles, 
the CIAO task {\it dmextract} has been applied to each dataset individually
with appropriate source and background apertures, and the results are 
weighted by the exposure time and averaged to produce the combined
count rate.

Firm, $\gtrsim3\sigma$ detections of X-ray emission are obtained for 26 WR 
stars in the LMC and 3 WR stars in the SMC, and tentative, $\sim2\sigma$ 
detections are found for four additional WR stars in the LMC.
Tables~3 and 4 list these firm and tentative detections, respectively.
The WR star identification is given in Columns 1 and 2, the offsets 
between the X-ray source and the position of the WR star in Columns 
3 and 4, and the instrument at the telescope aimpoint and detector 
used are given in Columns 5 and 6.  
The \emph{Chandra} X-ray images and accompanying optical images overlaid
with X-ray contours of these 33 WR stars with firm or tentative X-ray
detections are displayed in Figure~1.  
In a large number of WR stars, the spatial coincidence of the X-ray source 
and the WR star is within 1\arcsec, i.e., within the astrometry accuracy 
of the data, and in most cases the offset is within 2\arcsec.  
There are, however, a few excursions;  
notably, the X-ray sources associated to Brey\,16a, Brey\,95a (Tab.~3), and 
Brey\,74a (Tab.~4) are $\sim$2\farcs4, $\sim$4\farcs0, and $\sim$4\farcs8 
from the nominal positions of these WR stars.  
Brey\,74a is embedded within diffuse X-ray emission and the tentative 
detection of X-ray emission from this star may be questioned in view 
of the large offset.  
On the other hand, a systematic offset of 1\farcs4--1\farcs5 is found 
for other WR stars, namely Brey\,94 and Brey\,95, detected in the 
field of view of Brey\,95a.  
This suggests that there is a systematic error in the position of this 
\emph{Chandra} observation, and the same may also apply to the \emph{Chandra} 
observation of Brey\,16a, based on the observed offset of Brey\,16 also in 
this field of view.  
Therefore, we are confident of the X-ray detections of Brey\,16a and Brey\,95a 
in spite of the large offsets between these WR stars and the X-ray sources 
associated to them.

Columns 7, 8, and 9 of Tabs.~3 and 4 provide the net exposure, and 
the background-subtracted count rate and counts in Columns 8 and 9, 
respectively.  
The 1-$\sigma$ deviations given in Column 9 are calculated adding quadratically 
the background rms and the Poisson error of the source count number estimated 
using the approximation given by equation 7 of \citet{G86}.  
Basic spectral information is provided by Columns 10 and 11, that list 
the hardness ratio (defined as the number of counts in the 2.0--7.0 
keV band divided by the number of counts in the 0.3--7.0 keV band), 
and the median energy in the 0.3-7.0 keV band, respectively.

Some of the WR stars in Tables~3 and 4 have been reported previously
to be associated with X-ray sources.
HD\,5980 in the SMC is superposed on a supernova remnant;  
its stellar X-ray emission was resolved from the supernova 
remnant for the first time by the \emph{Chandra} observations 
reported by \citet{Netal02}.  
\citet{W95} used \emph{ROSAT} PSPC observations of the 30 Doradus nebula 
to study the spectral properties of the X-ray bright Brey\,84 and R\,140a.  
More recently, \citet{PPL02} and \citet{Tetal06} have analyzed a 
\emph{Chandra} observation of the same region and detected X-ray 
emission from 10-20 WR stars.  
Finally, \citet{FMG03a} and \citet{FMG03b} have reported preliminary 
results of the survey presented in this paper.  
Further comparisons with the results presented in these papers are 
made in \S3.2.1.

The WR stars that are not detected by \emph{Chandra} ACIS observations 
are listed in Table~5. 
The WR star identification is given in Columns 1 and 2, the instrument 
at the telescope aim point and detector used in Columns 3 and 4, the 
exposure time in Column 5, and the 3-$\sigma$ upper limit in Column 6, 
respectively.  
The 3-$\sigma$ upper limit is calculated from the rms of the local 
background, scaling it to the size of the source aperture.  
Three of these undetected WR stars, LMC-WR\,68 (Brey\,58), LMC-WR\,69 
(TSWR\,4), and LMC-WR\,115 (Brey\,83), are located close to bright 
X-ray sources, but outside their error radius. 
% of a few arcsec.
Similarly, LMC-WR\,111 (R\,136b) is located between two bright X-ray 
sources, R\,136a, at the center of the R\,136 star cluster R\,136, and
R\,136c, but no X-ray point source is detected at the location of R\,136b.

\subsection{Spectral Analysis of Individual WR Stars in the MCs}

The X-ray spectra of the WR stars in Tab.~3 and 4 can be used to 
determine the physical conditions of the X-ray-emitting gas and
the absorption column density ($N_{\rm H}$) of the intervening 
material, which in turn can help us understand the origin of the 
X-ray emission.  
Spectral analysis is achieved by modeling the observed spectra,  
convolving absorbed plasma emission models with the instrumental 
response, and using $\chi^2$ statistics to determine the model 
parameters that best fit an observed spectrum.  
Meaningful spectral analysis with $\chi^2$ statistics can be performed 
only when an adequate number of source counts are detected.  
For WR stars whose spectra have a low number of counts, a CSTAT 
fitting of an absorbed plasma emission model or of a spline fit 
can be performed to the unbinned spectrum to derive their observed 
luminosities.  

To carry out these different levels of spectral analysis for the WR 
stars detected in X-rays, we have divided these WR stars into two
subgroups: the high-count X-ray WR stars whose observations detected
at least 70 counts, and the low-count X-ray WR stars whose observations
detected fewer than 70 counts.
As we show later, the high-count WR stars are generally brighter than the 
low-count WR stars, but a few high-count WR stars have been observed with 
very long exposures and they are actually as X-ray faint as those in the 
low-count subsample.

\subsubsection{The Subsample of High-Count X-ray WR Stars in the MCs}

The background-subtracted X-ray spectra of WR stars in the high-count 
subsample are presented in Figure~2.
These spectra show that most of the counts are detected in the 
0.7--2.0 keV energy range, and that very few counts are detected
below 0.5 keV.
Some spectra show high-energy tails extending beyond 5 keV.
These spectral features are suggestive of high levels of absorption 
and high temperature X-ray-emitting gas ($kT\ge1$ keV).

To model the spectra of these stars, we have first created the response 
matrix and effective area files (the so-called redistribution matrix 
file, RMF, and auxiliary response file, ARF) associated with each spectrum.  
If a WR star has multiple short observations (e.g., SMC-WR\,6 
and SMC-WR\,7), we compute the RMF and ARF files individually 
for each observation and take the exposure-weighted average 
of them all.

Our selection of appropriate spectral models for plasma emission and
absorption along the line-of-sight is guided by the detailed spectral 
analysis of \emph{Chandra} and \emph{XMM-Newton} observations of 
Galactic WR stars available in the literature.
The X-ray spectra of the Galactic WR binaries WR\,25 and 
$\gamma^2$~Velorum can be described by thin plasma emission 
at temperatures $kT$ from 0.3--0.7 keV up to 2--3 keV 
\citep{Retal03,Setal01}.   
Similarly, the X-ray spectra of the presumably single Galactic WR stars 
WR\,110 and EZ CMa show both cool and hot plasma components, suggesting
either the presence of an unknown close stellar companion or the 
possibility that hard X-rays can be produced by single WR stars 
\citep{Setal02a,Setal02b}.  
We note that the observations of WR stars in the MCs do not detect as 
many X-ray photons as these of Galactic WR stars considered above.  
Moreover, after re-binning the spectra of WR stars in the MCs 
to $\ge$ 16 counts per energy bin, as required for a reliable 
spectral fit using the $\chi^2$ statistics, the spectral resolution 
becomes lower, too.  
Therefore, as the spectra of WR stars in the MCs have insufficient 
quality for modeling with multiple temperature components, we have 
adopted a single-temperature MEKAL optically thin plasma emission 
model \citep{KM93,LOG95}.
As for the absorption along the line-of-sight, we used the 
photoelectric absorption model of \citet{BM92}.  
Finally, the chemical abundances of the X-ray-emitting plasma and of the 
absorbing intervening material were fixed to be 0.33 $Z_\odot$ for the 
LMC, and 0.25 $Z_\odot$ for the SMC\footnote{ 
The X-ray-emitting material in the WR winds is expected to be 
highly enriched; however, detailed spectral fitting of 
high-dispersion X-ray spectra of the bright Galactic WR star 
WR\,25 did not find strong metal enrichment \citep{Retal03}.  
Given the limited quality of the X-ray spectra of the WR stars in the 
MCs, we have chosen to simply use the canonical abundances of the MCs.  
While our adopted abundances may potentially yield unreliable 
best-fit parameters, the derived X-ray luminosities are not 
very sensitive to the chemical abundances of X-ray-emitting and 
absorbing material.  
}.  
The absorption due to contamination build-up on the filters of 
the ACIS detectors is already taken into account by the ARF for
these datasets reprocessed with CIAO v3.1 and CALDB 2.28.  
For those other datasets obtained at a focal plane temperature 
different than $-$120$^\circ$C, this absorption has been modeled 
with the XSPEC ACISABS absorption model with the default 
composition for the contaminant and the amplitude of the 
absorption adjusted by the date of the observation\footnote{
Further information on the low energy quantum efficiency 
degradation of ACIS can be obtained at 
http://asc.harvard.edu/cal/Acis/Cal$_-$prods/qeDeg. }

Spectral fits were performed using XSPEC v11.0.1.  
The best-fit spectral models are over-plotted on the background-subtracted 
spectra shown in Figure~2.  
The best-fit model parameters ($kT$ and $N_{\rm H}$), reduced $\chi^2$ 
($\chi^2$ per degree of freedom), the observed flux in the 0.5-7.0 keV 
energy band, and the intrinsic luminosity in the total (0.5-7.0 keV) and 
soft (0.5-2.0 keV) energy bands are listed in Table~6.  
The best-fit model parameters quantitatively confirm the high plasma 
temperatures and large absorption columns suggested by the spectral 
shape.
The best-fit absorption column densities are much larger than 
those expected from the average extinction towards the MCs and 
most likely indicate large amounts of intrinsic extinction.  
We note that LMC-WR\,19, 20, and 99 have small numbers of counts 
detected, and consequently the quality of their spectral fits 
is poor.  
Spectral fits were not statistically improved by adding a second thin 
plasma emission component at a different temperature, except for SMC-WR\,6 
(Sk\,108) for which a soft component heavily extincted provided a formally 
better fit.  
This second component, however, increases dramatically the intrinsic 
X-ray luminosity of Sk\,108 and has been considered unrealistic.

It is worthwhile comparing the parameters of the best-fit models and 
X-ray luminosities listed in Tab.~6 with those derived by other authors.  
Table~7 compiles the relevant fit parameters and total unabsorbed 
X-ray luminosities of the WR stars in the MCs in common with 
\citet{Netal02}, \citet{PPL02}, and \citet{Tetal06}.  
\citet{Netal02} analyzed the X-ray spectrum of HD\,5980 (SMC-WR\,5) 
and derived a temperature and an absorption column density within 1 
$\sigma$ of our best-fit values; their X-ray luminosity of HD\,5980 is 
$\sim$50\% higher, but for a wider energy band, 0.3--10.0 keV.
In the 30 Doradus Nebula, \citet{PPL02} and \citet{Tetal06} have 
analyzed the X-ray spectra of Brey\,78, R140a, R136a, R136c, and 
Brey\,84.  
While the details of their spectral fits and ours are different, the 
best-fit parameters ($N_{\rm H}$ and $kT$) are within 1 $\sigma$ in 
all cases.  
Moreover, the total unabsorbed X-ray luminosities reported from the 
above studies are generally consistent with ours within a factor of 
$\lesssim$2.  
Differences in calibrations and in the details of the spectral fits (e.g., 
the energy band used for the spectral fit, the background region used, and 
the number of counts per spectral bin required for the spectral fit) and 
the use of chemical abundances of the LMCs for the intervening material 
may account for the detailed differences between our results and others.
We note that the results reported by \citet{Tetal06} and by \citet{PPL02} 
also show similar level of differences, and that our results are not 
systematically different from theirs.

\subsubsection{The Subsample of Low-Count X-ray WR Stars in the MCs}

For the MCs WR stars in the low-count subsample (i.e., those with 
fewer than 70 counts detected), individual spectral fitting using 
the $\chi^2$ statistics as for the WR stars in the high-count subsample 
is not feasible.  
A CSTAT fitting of a MEKAL absorbed thin-plasma emission model can 
still provide valuable information on the physical conditions of 
the X-ray-emitting plasma, although no goodness of the fit can be 
obtained.  
Spectral fits using the CSTAT statistic and a MEKAL absorbed thin-plasma 
emission model have been attempted, but the best-fit models resulted in 
unphysical values of the hydrogen column density.  
Furthermore, spectral fits could not be achieved in several of 
these WR stars.  
In these cases, a spline fit can be used to model the spectral shape 
and compute at least the X-ray flux, but the intervening column density 
of hydrogen is also needed in these cases to derive the intrinsic X-ray 
luminosity.

In order to fit the X-ray spectra of the low-count subsample of WR stars 
in the MCs, we need to adopt values of the intervening absorption
column density.
To derive an average absorption column density, we have respectively 
combined the BI and FI spectra of all MCs WR stars in the low-count 
subsample and performed a joint spectral fit of these two spectra using
a common absorbed MEKAL thin-plasma emission model and calibration
matrices obtained by adding the calibration matrices of each 
individual spectrum weighted by their background-subtracted counts.  
The best-fit model (Figure~3), with a plasma temperature of $kT$=1.6 keV, 
has an absorption column density of $\sim$3$\times$10$^{21}$~cm$^{-2}$ 
that will be adopted for spectral fits using either a MEKAL thin-plasma 
emission model or a spline function.

Table~8 lists the best-fit parameters, observed X-ray flux, and 
intrinsic X-ray luminosity in the 0.5-7.0 keV ($L_{\rm t}$) and 
0.5-2.0 keV ($L_{\rm s}$) energy bands as defined for Tab.~6.  
When a spline function was used to fit the spectral shape, no 
best-fit temperature is provided.  
It is interesting to compare the X-ray luminosities of the high-count 
and low-count WR stars in the MCs to search for systematic
differences in the emission models describing the spectral properties 
of these two samples.  
Figure~4 plots the intrinsic X-ray luminosity against the count 
rate for these two subsamples of WR stars.  
It is apparent that WR stars in the low-count subsample follow the 
same trend shown by the high-count subsample of WR stars to have 
higher X-ray luminosities for higher count rates, although they show a 
larger dispersion around the linear relationship.  
Figure~4 also shows that Brey\,16, Brey\,67, and R\,140a are much 
more luminous than expected from a linear relationship between 
X-ray luminosity and count rate.
These large discrepancies are caused by the large absorption 
column toward Brey\,16 and low plasma temperatures for Brey\,67 
and R\,140a, for which a greater correction from observed flux 
(i.e., count rate) to unabsorbed flux is required.

\subsubsection{The Undetected X-ray WR Stars in the MCs}

WR stars in the MCs that have been observed but not detected by 
\emph{Chandra} ACIS are listed in Tab.~5, together with the 3-$\sigma$ 
upper limits of their X-ray count rates.  
To further investigate the X-ray faintest population of WR stars 
in the MCs, we have stacked together all the spectra of the 
undetected WR stars and obtained combined BI and FI spectra 
(Figure~5).  
There is a clear detection of X-ray emission in the FI 
spectrum of 90$\pm$20 counts, with a total count rate 
of (1.4$\pm$0.3)$\times$10$^{-4}$ cnts~s$^{-1}$ for a 
total exposure time of 673.1~ks.  
In the BI spectrum, with a shorter integration time of 183.8 ks, the 
detection is unclear, although the spectral shape is tantalizing.  
We have then performed a joint spectral fit of these two spectra using 
a common absorbed MEKAL thin-plasma emission model and calibration matrices 
obtained by adding the calibration matrices of each individual spectrum 
weighted by their net exposure times.  
The spectral fit could not constrain the absorption column density, so
we adopted 3$\times$10$^{21}$~cm$^{-2}$, the averaged absorption column 
density derived for the low-count sources in $\S$3.2.2.  
The best-fit model, over-plotted in Fig.~5, has a two component plasma 
with temperatures $kT_1$=0.4 keV and $kT_2$=3 keV.  
The mean intrinsic luminosity in the energy bands 0.5-7.0 keV 
and 0.5-2.0 keV are 1.5$\times$10$^{31}$~ergs~s$^{-1}$ and 
1.1$\times$10$^{31}$~ergs~s$^{-1}$, respectively.

\subsection{Count Rate and Intrinsic X-ray Luminosity Distributions}

The histograms in Figure~6 show the distributions of the count rate (top) 
and intrinsic X-ray luminosity (bottom) for WR stars in the MCs that have 
been detected by \emph{Chandra} ACIS observations.  
The solid and dotted curves represent the high-count and low-count 
subsamples of WR stars in the MCs, respectively.  
WR stars in the high-count subsample are generally more luminous 
than the WR stars in the low-count subsample, although there is 
some overlap.  
We do not detect any source with a count rate lower than 
$\simeq$1$\times$10$^{-4}$ cnts~s$^{-1}$, or 
$\simeq$6$\times$10$^{32}$ ergs~s$^{-1}$ in X-ray luminosity.
The vertical dashed line shown in the plot of count rate distributions 
marks the average count rate of the undetected sources derived in the 
previous section.

\subsection{X-ray Variability}

X-ray variability can be investigated for WR stars detected with 
large numbers of counts.  
Note that count rates for multiple observations of a WR star made with 
different detectors or with the star at different offsets from the 
telescope aimpoint cannot be compared directly.  
Instead, we compared photon flux, in units of photons~cm$^{-2}$~s$^{-1}$,
derived from each observation by dividing the number of source counts by 
the effective exposure time and the aperture area.
To analyze the X-ray variability of the whole sample of sources in a 
homogeneous manner, we use photon fluxes also for those with a single, 
long observation for which count rates could have been used directly 
for light curve analysis.
The light curves of the WR stars with sufficient number of counts and 
time coverage are shown in Figure~7.  
The photon fluxes of the WR stars with low number of counts or few 
time points are listed in Table~9.

The light curves of Brey\,56, Brey\,78, R\,140a, R\,136a, and Sk\,108 
in Fig.~7, and the photon fluxes at different epochs of Brey\,65b and 
Brey\,67 in Tab.~9 are consistent with a constant level of X-ray 
emission.
On the other hand, the light curves of Brey\,84 and R\,136c in the LMC, and 
AV\,336a and HD\,5980 in the SMC show small variations that are indicative 
of X-ray variability:  
Brey\,84 presents a modulation in its light curve with a period of 
$\sim$2 hours, as revealed by the task \emph{efsearch} of the HEASOFT 
XRONOS package;  
R\,136c light curve shows similar flux drops separated by $\sim$5 hours;
AV\,336a light curve shows an apparent modulation consistent with the 
orbital period of this binary system, 19.560 days \citep{NMetal02}; and  
HD\,5980's count rate progressively increased by a factor of two during
its observation \citep{Netal02}.  
Furthermore, the count rates of Brey\,16 and Brey\,16a listed in Tab.~9 
varied significantly from 1999 December to 2003 February: 
the photon flux of Brey\,16 increased by a factor 2, while that of 
Brey\,16a decreased by about the same factor in 2003.

\subsection{Individual Objects}

{\bf Brey\,16} (LMC-WR\,19) is a WN4b+O5: spectroscopic binary \citep{SSM96} 
that was recently revealed to be an eclipsing binary \citep{FMG03a}.  
The double eclipse seen in its optical light curve and the sinusoidal 
variations of its radial velocity indicate an orbital period of $\sim$18.0 
days.  
The short orbital period makes Brey\,16 a good candidate for a 
colliding-wind binary.  
Indeed, the hardness of the X-ray spectrum of LMC-WR\,19 and the high 
temperature of the plasma emission are consistent with the emission 
expected from a colliding-wind binary \citep{PS97}.  
Moreover, the \emph{Chandra} observations at two different epochs show a 
photon flux increase by a factor 2 that suggests an X-ray variability of 
this system that can be associated with the orbital phase.  
Using the orbital parameters of Brey\,16 derived by \citet{FMG03a}, we 
find that \emph{Chandra}'s first epoch observation spanned from orbital 
phase $\sim$0.06 to $\sim$0.075, i.e., it sampled the start of the 
secondary eclipse shown in Brey\,16 light curve, while \emph{Chandra}'s 
second epoch observation occurred at orbital phase 0.40--0.42, at an 
intermediate phase between the two eclipses.  
Therefore, there is evidence that the X-ray variability of Brey\,16 is 
phase-locked: the passage of one star in front of the other during the 
secondary eclipse also implies the absorption of a significant fraction 
of the X-ray emission from the colliding-wind zone.

{\bf R\,140a} (LMC-WR\,101,102) and {\bf R\,140b} (LMC-WR\,103) are the WR 
components of the visual multiple system R\,140 in the 30 Doradus region.  
R\,140a1 (LMC-WR\,101) is a WC5+O(?) binary system, and R\,140a2 
(LMC-WR\,102) is a WN6+O binary system with an orbital period of $\sim$2.76 
days \citep{M89}, while R\,140b is not known to be binary \citep{BAT99}.  
R\,140a (not resolved by \emph{Chandra} ACIS) is one of the 
brightest X-ray sources in the 30 Doradus region, while 
R\,140b is only tentatively detected with $\sim$7 counts.  
It must be noted that \citet{Tetal06} reported a firm detection of 
R\,140b with $\sim$15 counts.  
It is unclear the origin of this discrepancy, but the proximity of R\,140b to 
the bright X-ray source R\,140a made us use a conservatively small aperture 
that may have miss a fraction of its flux, while \citet{Tetal06} refined the 
event positions using a sub-pixel positioning technique that is best suited in 
order to select photons from a faint source nearby a bright one.

{\bf R\,136a} (LMC-WR\,106,108-110), {\bf R\,136b} (LMC-WR\,111), and 
{\bf R\,136c} (LMC-WR\,112) are the WR components of the central cluster 
of 30 Doradus.  
R\,136a is a bright X-ray source and there is evidence for X-ray 
variability with a short period of $\sim$5 hours.  
The detailed spatial analysis of this region by \citet{Tetal06} was able to 
resolve it into several of its components.  
Similarly, the improved spatial resolution of \citet{Tetal06} images made 
possible the detection of R\,136b, too nearby to the X-ray bright R\,136a 
to be resolved by us.

{\bf Brey\,84} (LMC-WR\,116) is the brightest X-ray source in the 30 Doradus 
region and, due to its high X-ray luminosity, its true nature has been 
disputed.  
Based on \emph{ROSAT} observations, \citet{W95} proposed it to be a 
high-mass X-ray binary composed of a WR star and a black hole.  
The \emph{Chandra} ACIS X-ray spectrum, however, can be fitted with a 
thin-plasma emission model whose temperature and X-ray luminosity are 
typical of colliding-wind X-ray sources, although at the high end of 
the distribution of observed parameters \citep{PPL02}.  
While the X-ray properties of Brey\,84 do not require the presence of
a black hole, the hard X-ray spectrum, high X-ray luminosity, and the 
$\sim$2 hr modulation of the X-ray light curve are intriguing, and further
investigation is warranted.

{\bf HD\,5980} (SMC-WR\,5) in the SMC is a massive binary system in which 
one component is a WR star and the other is in a Luminous Blue Variable 
(LBV) phase evolving towards a WR star \citep{K04}.  
The orbital period is short, $\sim$19.266 days \citep{BP80}, and the 
stellar winds of both components should be able to generate strong 
shocks in the colliding-wind zone.  
The X-ray properties of this system are described in detail by 
\citet{Netal02}.

{\bf AV\,336a} (SMC-WR\,7) is a WN+O6 spectroscopic binary with an orbital 
period of 19.560 days \citep{NMetal02}.  
When the X-ray light curve of AV\,336a is folded with its orbital period 
(Fig.~7), periodic variations are observed in the light curve.

\section{Summary and Future Work}

We have analyzed all serendipitous \emph{Chandra} ACIS observations of 
WR stars in the MCs available up to October 2004 in the \emph{Chandra} 
archive.  
This survey for X-ray emission from WR stars in the MCs has included 192 
archival observations for 61 WR stars in the LMC and 9 WR stars in the SMC,
i.e., about half of the known WR stars in the MCs.  
X-ray emission is confidently detected from 29 WR stars and possibly 
4 more WR stars, rendering a detection rate of $\sim$50\% among the 
WR stars in the MCs.  
The faintest WR star detected in this survey has an intrinsic X-ray 
luminosity of 6$\times$10$^{32}$ ergs~s$^{-1}$.

We have modeled the X-ray spectra of the 12 WR stars that have 
sufficient number of counts for spectral analysis.
The majority of the best-fit models show high plasma temperatures
and large absorption column densities.
Similar X-ray spectral behavior is exhibited in the integrated
spectrum of the WR stars detected with small number of counts.

The X-ray variability of WR stars in the MCs has also been investigated.
R\,136c and Brey\,84, among the X-ray brightest WR stars in the LMC, 
show evidence of short-term variability with periods of $\sim$5 hours 
and $\sim$2 hours, respectively.  
Brey\,16 and Brey\,16a in the LMC also show a factor of 2 variations 
in their X-ray emission between two observations separated by 3.2 yr.
The variations observed in Brey\,16 are consistent with the orbital 
period of $\sim$18 days of this binary system, thus suggesting that 
its X-ray variability is phase-locked.  
Similarly, the variations observed in the X-ray light curve of 
the binary system AV\,336a in the SMC are consistent with its 
orbital period of 19.560 days.  
Finally, HD\,5980 in the SMC also appears to vary.

In subsequent papers, we will use X-ray observations of WR stars in the MCs 
available in the \emph{ROSAT} and \emph{XMM-Newton} archives to acquire the 
most comprehensive view of X-ray emission from WR stars in the MCs.  
This work will be used to investigate the occurrence, luminosity 
and variability of X-ray emission in WR stars with their spectral 
type and binarity properties.  
The final results will help us determine the X-ray properties of the WR 
stars in the MCs, and allow comparisons with their Galactic counterparts.

\acknowledgments

This work is supported by the \emph{Chandra X-ray Observatory} 
grant AR3-4001X.  
M.A.G.\ also acknowledges support from the grants AYA 2002-00376
and AYA 2005-01495 of the Spanish MEC (co-funded by FEDER funds)
and the Spanish program Ram\'on y Cajal.

\clearpage

{\tiny 
\begin{deluxetable}{llrrrllrr}
\tablenum{1}
\tablewidth{0pc}
\tabletypesize{\scriptsize} 
\tablecaption{Chandra ACIS Observations of WR Stars in the LMC}
\tablehead{
\multicolumn{1}{c}{WR $\#$} & 
\multicolumn{1}{c}{WR Name} & 
\multicolumn{1}{c}{$\alpha$} & 
\multicolumn{1}{c}{$\delta$} & 
\multicolumn{1}{c}{Obs.\ ID} & 
\multicolumn{1}{c}{Aimpoint} & 
\multicolumn{1}{c}{CCD} & 
\multicolumn{1}{c}{$t_{\rm exp}^{\rm obs}$} & 
\multicolumn{1}{c}{$t_{\rm exp}$} \\
\multicolumn{1}{c}{} & 
\multicolumn{1}{c}{} & 
\multicolumn{2}{c}{(J2000)} & 
\multicolumn{1}{c}{} & 
\multicolumn{1}{c}{} & 
\multicolumn{1}{c}{} & 
\multicolumn{1}{c}{(ks)} & 
\multicolumn{1}{c}{(ks)} 
}

\startdata 
 LMC-WR\,19 & Brey\,16          & 05:09:40.45 & --68:53:24.3 &   125 & ACIS-S & I2 & 36.7 & 30.4 \\
            &                   &             &              &  3810 & ACIS-S & S2 & 29.7 & 29.6 \\
 LMC-WR\,20 & Brey\,16a         & 05:09:53.85 & --68:52:51.6 &   125 & ACIS-S & I2 & 36.7 & 30.4 \\
            &                   &             &              &  3810 & ACIS-S & S2 & 29.7 & 29.6 \\
 LMC-WR\,31 & Brey\,25          & 05:22:04.34 & --67:59:06.2 &  3356 & ACIS-S & S3 & 18.7 & 18.6 \\
 LMC-WR\,49 & Brey\,40a         & 05:29:33.16 & --70:59:35.2 &  3848 & ACIS-S & I3 & 33.1 & 33.1 \\
            &                   &             &              &  4421 & ACIS-S & I3 & 34.8 & 34.7 \\
 LMC-WR\,60 & Brey\,49          & 05:33:10.74 & --69:29:00.3 &   122 & ACIS-I & I1 &  8.6 &  8.4 \\
\enddata
\end{deluxetable}
}

\clearpage

{\tiny 
\begin{deluxetable}{llrrrllrr}
\tablenum{2}
\tablewidth{0pc}
\tabletypesize{\scriptsize} 
\tablecaption{Chandra ACIS Observations of WR Stars in the SMC}
\tablehead{
\multicolumn{1}{c}{WR $\#$} & 
\multicolumn{1}{c}{WR Name} & 
\multicolumn{1}{c}{$\alpha$} & 
\multicolumn{1}{c}{$\delta$} & 
\multicolumn{1}{c}{Obs.\ ID} & 
\multicolumn{1}{c}{Aimpoint} & 
\multicolumn{1}{c}{CCD} & 
\multicolumn{1}{c}{$t_{\rm exp}^{\rm obs}$} & 
\multicolumn{1}{c}{$t_{\rm exp}$} \\
\multicolumn{1}{c}{} & 
\multicolumn{1}{c}{} & 
\multicolumn{2}{c}{(J2000)} & 
\multicolumn{1}{c}{} & 
\multicolumn{1}{c}{} & 
\multicolumn{1}{c}{} & 
\multicolumn{1}{c}{(ks)} & 
\multicolumn{1}{c}{(ks)} 
}

\startdata 
SMC-WR\,2   & AV\,39a            & 00:48:30.81 & --73:15:45.1 &  2945 & ACIS-I & I3 & 11.6 & 11.6 \\
SMC-WR\,3   & AV\,60a            & 00:49:59.33 & --73:22:13.6 &  2945 & ACIS-I & I0 & 11.6 & 11.6 \\
SMC-WR\,5   & HD\,5980           & 00:59:26.60 & --72:09:53.5 &   444 & ACIS-I & S2 &  8.4 &  8.2 \\
            &                    &             &              &  1543 & ACIS-I & S2 &  7.4 &  6.5 \\
            &                    &             &              &  1881 & ACIS-I & I1 & 98.7 & 98.5 \\
            &                    &             &              &  2862 & ACIS-I & S2 &  7.6 &  7.5 \\
            &                    &             &              &  5137 & ACIS-I & S2 &  8.0 &  8.0 \\
SMC-WR\,6   & Sk\,108            & 01:03:25.20 & --72:06:43.6 &   48 & ACIS-I & I2 &  8.6 &  8.4 \\
\enddata
\end{deluxetable}
}

\clearpage

{\tiny 
\begin{deluxetable}{llrrllrrrcr}
\tablenum{3}
\tablewidth{0pc}
\tabletypesize{\scriptsize} 
\tablecaption{WR Stars in the MCs with Detected X-ray Emission}
\tablehead{
\multicolumn{1}{c}{WR $\#$} & 
\multicolumn{1}{c}{WR Name} & 
\multicolumn{1}{c}{$\Delta\alpha$} & 
\multicolumn{1}{c}{$\Delta\delta$} &
\multicolumn{1}{c}{Aimpoint} & 
\multicolumn{1}{c}{CCD} & 
\multicolumn{1}{c}{$t_{\rm exp}$} & 
\multicolumn{1}{c}{Count Rate} & 
\multicolumn{1}{c}{Counts} & 
\multicolumn{1}{c}{Hardness Ratio} & 
\multicolumn{1}{c}{Median Energy} \\
\multicolumn{1}{c}{} &
\multicolumn{1}{c}{} &
\multicolumn{1}{c}{($\prime\prime$)} &
\multicolumn{1}{c}{($\prime\prime$)} &
\multicolumn{1}{c}{} &
\multicolumn{1}{c}{} &
\multicolumn{1}{c}{(ks)} &
\multicolumn{1}{c}{(cnts~s$^{-1}$)} &
\multicolumn{1}{c}{(cnts)} &
\multicolumn{1}{c}{$\frac{C_{2.0-7.0 {\rm keV}}}{C_{0.3-7.0 {\rm keV}}}$} & 
\multicolumn{1}{c}{(keV)} 

}

\startdata
LMC-WR\,19  & Brey\,16   & $-$0.79~ & $+$0.11~ & ACIS-S & I2    &  30.4 & 1.3$\times$10$^{-3}$~ &  40$\pm$7~ & 0.71 & 2.5~~~~~~~~ \\
            &            &          &          & ACIS-S & S2    &  29.6 & 2.4$\times$10$^{-3}$~ &  70$\pm$10 & 0.56 & 2.3~~~~~~~~ \\
LMC-WR\,20  & Brey\,16a  & $-$2.39~ & $-$0.25~ & ACIS-S & I2    &  30.4 & 1.7$\times$10$^{-3}$~ &  51$\pm$8~ & 0.25 & 1.5~~~~~~~~ \\
            &            &          &          & ACIS-S & S2    &  29.6 & 6.8$\times$10$^{-4}$~ &  20$\pm$7~ & 0.43 & 1.6~~~~~~~~ \\
LMC-WR\,67  & Brey\,56   & $+$0.07~ & $-$0.66~ & ACIS-S & S3    & 271.2 & 3.6$\times$10$^{-4}$~ &  97$\pm$14 & 0.29 & 1.6~~~~~~~~ \\
LMC-WR\,78  & Brey\,65b  & $-$0.52~ & $-$1.21~ & ACIS-S & S2    &  91.7 & 1.2$\times$10$^{-4}$~ &  11$\pm$3~ & 0.27 & 1.6~~~~~~~~ \\
            &            &          &          & ACIS-S & S3    &  58.0 & 1.7$\times$10$^{-4}$~ &  10$\pm$6~ & 0.43 & 1.1~~~~~~~~ \\ 
LMC-WR\,79  & Brey\,57   & $-$0.48~ & $-$1.79~ & ACIS-S & S3    &  58.0 & 3.1$\times$10$^{-4}$~ &  19$\pm$7~ & 0.02 & 1.2~~~~~~~~ \\
\enddata
\end{deluxetable}
}

\clearpage

{\tiny 
\begin{deluxetable}{llrrllrrrcr}
\tablenum{4}
\tablewidth{0pc}
\tabletypesize{\scriptsize} 
\tablecaption{WR Stars in the MCs with Tentatively Detected X-ray Emission}
\tablehead{
\multicolumn{1}{c}{WR $\#$} & 
\multicolumn{1}{c}{WR Name} & 
\multicolumn{1}{c}{$\Delta\alpha$} & 
\multicolumn{1}{c}{$\Delta\delta$} &
\multicolumn{1}{c}{Aimpoint} & 
\multicolumn{1}{c}{CCD} & 
\multicolumn{1}{c}{$t_{\rm exp}$} & 
\multicolumn{1}{c}{Count Rate} & 
\multicolumn{1}{c}{Counts} & 
\multicolumn{1}{c}{Hardness Ratio} & 
\multicolumn{1}{c}{Median Energy} \\
\multicolumn{1}{c}{} &
\multicolumn{1}{c}{} &
\multicolumn{1}{c}{($\prime\prime$)} &
\multicolumn{1}{c}{($\prime\prime$)} &
\multicolumn{1}{c}{} &
\multicolumn{1}{c}{} &
\multicolumn{1}{c}{(ks)} &
\multicolumn{1}{c}{(cnts~s$^{-1}$)} &
\multicolumn{1}{c}{(cnts)} &
\multicolumn{1}{c}{$\frac{C_{2.0-7.0 {\rm keV}}}{C_{0.3-7.0 {\rm keV}}}$} & 
\multicolumn{1}{c}{(keV)}
}

\startdata
LMC-WR\,77  & Brey\,65   & $+$0.07~ & $-$0.66~ & ACIS-S & S3       &  58.0 & 3.1$\times$10$^{-4}$~ &  18$\pm$7~ & 0.35 & 1.7~~~~~~~~ \\ 
LMC-WR\,82  & Brey\,66   & $-$0.05~ & $-$0.66~ & ACIS-S & I3,S1,S4 &  61.4 & 2.6$\times$10$^{-4}$~ &  16$\pm$7~ & 0.00 & 1.0~~~~~~~~ \\ 
LMC-WR\,93  & Brey\,74a  & $+$4.71~ & $-$1.15~ & ACIS-S & S3       &  48.2 & 5.0$\times$10$^{-4}$~ &  24$\pm$9~ & 0.22 & 1.1~~~~~~~~ \\ 
LMC-WR\,103 & R140b      & $-$0.11~ & $+$0.06~ & ACIS-I & I2       &  17.6 & 3.8$\times$10$^{-4}$~ &   7$\pm$3~ & 0.44 & 1.3~~~~~~~~ 
\enddata
\end{deluxetable}
}

\clearpage
  
{\tiny 
\begin{deluxetable}{llllrrcl}
\tablenum{5}
\tablewidth{0pc}
\tabletypesize{\scriptsize} 
\tablecaption{WR Stars in the MCs Undetected in X-rays}
\tablehead{
\multicolumn{1}{c}{WR $\#$} & 
\multicolumn{1}{c}{WR Name} & 
\multicolumn{1}{c}{Aimpoint} & 
\multicolumn{1}{c}{CCD} & 
\multicolumn{1}{c}{$t_{\rm exp}$} & 
\multicolumn{1}{c}{3-$\sigma$ Upper Limit}  \\
\multicolumn{1}{c}{} &
\multicolumn{1}{c}{} &
\multicolumn{1}{c}{} &
\multicolumn{1}{c}{} &
\multicolumn{1}{c}{(ks)} &
\multicolumn{1}{c}{(cnts~s$^{-1}$)} 
}

\startdata
LMC-WR\,31  & Brey\,25   & ACIS-S & S3    &  18.6 & 1.1$\times$10$^{-3}$ \\
LMC-WR\,49  & Brey\,40a  & ACIS-S & I3    &  67.8 & 3.6$\times$10$^{-4}$ \\ 
LMC-WR\,60  & Brey\,49   & ACIS-I & I1    &   8.4 & 9.5$\times$10$^{-4}$ \\
            &            & ACIS-S & S1    &  15.9 & 1.1$\times$10$^{-3}$ \\
LMC-WR\,61  & Brey\,50   & ACIS-S & S2    &  59.4 & 3.0$\times$10$^{-4}$ \\ 
LMC-WR\,64  & Brey\,53   & ACIS-S & I1    &  59.2 & 2.3$\times$10$^{-4}$ \\ 
LMC-WR\,65  & Brey\,55   & ACIS-I & S3    &  15.5 & 9.0$\times$10$^{-4}$ \\
            &            & ACIS-S & I2,I3 & 105.7 & 2.0$\times$10$^{-4}$ \\
\enddata 
\end{deluxetable}
}

\clearpage

{\tiny 
\begin{deluxetable}{llcrccccc}
\tablenum{6}
\tablewidth{0pc}
%\rotate
\tabletypesize{\scriptsize} 
\tablecaption{Best-fit Parameters of High-Count X-ray WR Stars in the MCs}
\tablehead{
\multicolumn{1}{l}{WR $\#$} & 
\multicolumn{1}{l}{WR Name} & 
\multicolumn{1}{c}{Counts} & 
\multicolumn{1}{c}{$\chi^2$/DoF} & 
\multicolumn{1}{c}{$kT$} & 
\multicolumn{1}{c}{$N_{\rm H}$} & 
\multicolumn{1}{c}{$f_{\rm t}$} & 
\multicolumn{1}{c}{$L_{\rm t}$} & 
\multicolumn{1}{c}{$L_{\rm s}$} \\
\multicolumn{1}{c}{} &
\multicolumn{1}{c}{} &
\multicolumn{1}{c}{(cnts)} &
\multicolumn{1}{c}{} &
\multicolumn{1}{c}{(keV)} &
\multicolumn{1}{c}{(cm$^{-2}$)} & 
\multicolumn{1}{c}{(ergs~cm$^{-2}$~s$^{-1}$)} &  
\multicolumn{1}{c}{(ergs~s$^{-1}$)} & 
\multicolumn{1}{c}{(ergs~s$^{-1}$)}   
}

\startdata
LMC-WR\,19           & Brey\,16  & 110 &  3.1/11=0.3 & 7$^{+15}_{-5}$      & (3$^{+3}_{-1.5}$)$\times$10$^{22}$     & 6.1$\times$10$^{-14}$ & 3.8$\times$10$^{34}$ & 1.4$\times$10$^{34}$ \\
LMC-WR\,20           & Brey\,16a &  71 &  5.6/ 5=1.1 & 3$^{+9}_{-2}$       & (8$^{+20}_{-8}$)$\times$10$^{21}$      & 1.7$\times$10$^{-14}$ & 9.1$\times$10$^{33}$ & 4.7$\times$10$^{33}$ \\
LMC-WR\,67           & Brey\,56  &  97 &  8.7/13=0.7 & 2.3$^{+4}_{-1.2}$   & (1.4$^{+2.2}_{-0.9}$)$\times$10$^{22}$ & 2.3$\times$10$^{-15}$ & 1.7$\times$10$^{33}$ & 1.0$\times$10$^{33}$ \\
LMC-WR\,85           & Brey\,67  & 147 & 13.5/16=0.8 & 1.0$^{+0.4}_{-0.6}$ & (2.0$^{+1.4}_{-0.8}$)$\times$10$^{22}$ & 4.9$\times$10$^{-14}$ & 6.3$\times$10$^{34}$ & 5.6$\times$10$^{34}$ \\
LMC-WR\,99           & Brey\,78  &  77 &  1.0/ 1=1.0 & 1$^{+2}_{-1}$       & (4$^{+8}_{-4}$)$\times$10$^{21}$       & 2.4$\times$10$^{-14}$ & 1.7$\times$10$^{34}$ & 1.5$\times$10$^{34}$ \\
LMC-WR\,101,102      & R140a     & 330 & 31.0/17=1.8 & 0.9$\pm$0.2         & (9$^{+2}_{-2}$)$\times$10$^{21}$       & 9.6$\times$10$^{-14}$ & 1.6$\times$10$^{35}$ & 1.5$\times$10$^{35}$ \\
LMC-WR\,106,108--110 & R136a     & 164 &  8.5/13=0.7 & 2.1$^{+0.8}_{-0.5}$ & (1.9$^{+2.5}$)$\times$10$^{21}$        & 6.8$\times$10$^{-14}$ & 2.4$\times$10$^{34}$ & 1.5$\times$10$^{34}$ \\
LMC-WR\,112          & R136c     & 330 & 19.8/16=1.2 & 2.9$^{+1.5}_{-0.8}$ & (2.8$^{+3.0}_{-2.3}$)$\times$10$^{21}$ & 1.5$\times$10$^{-13}$ & 6.2$\times$10$^{34}$ & 3.3$\times$10$^{34}$ \\
\enddata
\end{deluxetable}
}

\clearpage

{\tiny 
\begin{deluxetable}{llrcccl}
\tablenum{7}
\tablewidth{0pc}
%\rotate
\tabletypesize{\scriptsize} 
\tablecaption{Comparison of Spectral Fits of WR Stars in the MCs}
\tablehead{
\multicolumn{1}{l}{WR $\#$} & 
\multicolumn{1}{l}{WR Name} & 
\multicolumn{1}{c}{$\chi^2$/DoF} & 
\multicolumn{1}{c}{$kT$} & 
\multicolumn{1}{c}{$N_{\rm H}$} & 
\multicolumn{1}{c}{$L_{\rm t}$} & 
\multicolumn{1}{l}{Reference} \\
\multicolumn{1}{c}{} &
\multicolumn{1}{c}{} &
\multicolumn{1}{c}{} &
\multicolumn{1}{c}{(keV)} &
\multicolumn{1}{c}{(cm$^{-2}$)} & 
\multicolumn{1}{c}{(ergs~s$^{-1}$)} & 
\multicolumn{1}{c}{}   
}

\startdata
LMC-WR\,99           & Brey\,78  &  1.0/ 1=1.0 & 1$^{+2}_{-1}$         & (4$^{+8}_{-4}$)$\times$10$^{21}$       & 1.7$\times$10$^{34}$ & This paper \\
                     &           &  8.0/ 6=1.3 & 2.1                   & 1.6$\times$10$^{21}$                   & 1.0$\times$10$^{34}$ & \citet{PPL02} \\
                     &           & $\dots$     & 2.0$^{+2.6}$          & (1.6$^{+2.4}$)$\times$10$^{21}$        & 1.7$\times$10$^{34}$ & \citet{Tetal06} \\
\hline 
LMC-WR\,101,102      & R140a     & 31.0/17=1.8 & 0.9$\pm$0.2           & (9$^{+2}_{-2}$)$\times$10$^{21}$       & 1.6$\times$10$^{35}$ & This paper \\
                     &           & 11.4/17=0.7 & 1.1                   & 7.9$\times$10$^{21}$                   & 1.7$\times$10$^{35}$ & \citet{PPL02} \\
                     &           & $\dots$     & 0.9$^{+0.10}_{-0.11}$ & (7.9$^{+1.6}_{-1.5}$)$\times$10$^{21}$ & 1.8$\times$10$^{35}$ & \citet{Tetal06} \\
\hline 
LMC-WR\,106,108--110 & R136a     &  8.5/13=0.7 & 2.1$^{+0.8}_{-0.5}$   & (1.9$^{+2.5}$)$\times$10$^{21}$         & 2.4$\times$10$^{34}$ & This paper \\ % !!
                     &           & 14.6/13=1.1 & 1.7                   & 2.8$\times$10$^{21}$                   & 2.2$\times$10$^{34}$ & \citet{PPL02} \\
\enddata
\end{deluxetable}
}

\clearpage

{\tiny 
\begin{deluxetable}{llrccccc}
\tablenum{8}
\tablewidth{0pc}
%\rotate
\tabletypesize{\scriptsize} 
\tablecaption{Best-fit Parameters of Low-Count X-ray WR Stars in the MCs}
\tablehead{
\multicolumn{1}{l}{WR $\#$} & 
\multicolumn{1}{l}{WR Name} & 
\multicolumn{1}{c}{Counts} & 
\multicolumn{1}{c}{Spectral Fit} & 
\multicolumn{1}{c}{$kT$} &
\multicolumn{1}{c}{$f_{\rm t}$} & 
\multicolumn{1}{c}{$L_{\rm t}$} & 
\multicolumn{1}{c}{$L_{\rm s}$} \\
\multicolumn{1}{c}{} &
\multicolumn{1}{c}{} &
\multicolumn{1}{c}{(cnts)} &
\multicolumn{1}{c}{} &
\multicolumn{1}{c}{(keV)} & 
\multicolumn{1}{c}{(ergs~cm$^{-2}$~s$^{-1}$)} &  
\multicolumn{1}{c}{(ergs~s$^{-1}$)} & 
\multicolumn{1}{c}{(ergs~s$^{-1}$)}   
}

\startdata

LMC-WR\,77  & Brey\,65  & 18~~ & SPLINE & $\dots$ & 3.3$\times$10$^{-15}$ & 1.4$\times$10$^{33}$ & 6.4$\times$10$^{32}$ \\
LMC-WR\,78  & Brey\,65b & 21~~ & SPLINE & $\dots$ & 2.0$\times$10$^{-15}$ & 6.8$\times$10$^{32}$ & 2.3$\times$10$^{32}$ \\
LMC-WR\,79  & Brey\,57  & 19~~ & MEKAL  & 1.1     & 1.7$\times$10$^{-15}$ & 7.4$\times$10$^{32}$ & 6.2$\times$10$^{32}$ \\
LMC-WR\,80  & Brey\,65c & 77~~ & MEKAL  & 0.6,8.6 & 2.5$\times$10$^{-15}$ & 1.0$\times$10$^{33}$ & 6.9$\times$10$^{32}$ \\
LMC-WR\,82  & Brey\,66  & 16~~ & MEKAL  & 0.5     & 2.8$\times$10$^{-15}$ & 1.6$\times$10$^{33}$ & 1.5$\times$10$^{33}$ \\
LMC-WR\,92  & Brey\,72  & 12~~ & MEKAL  & 1.5     & 9.9$\times$10$^{-15}$ & 4.5$\times$10$^{33}$ & 2.7$\times$10$^{33}$ \\
LMC-WR\,93  & Brey\,74a & 24~~ & MEKAL  & 0.9     & 1.8$\times$10$^{-15}$ & 7.9$\times$10$^{32}$ & 2.7$\times$10$^{32}$ \\
LMC-WR\,100 & Brey\,75  &  8~~ & SPLINE & $\dots$ & 4.5$\times$10$^{-15}$ & 1.7$\times$10$^{33}$ & 1.0$\times$10$^{33}$ \\
\enddata
\end{deluxetable}
}

\clearpage

{\tiny 
\begin{deluxetable}{llccrccr}
\tablenum{9}
\tablewidth{0pc}
\tabletypesize{\scriptsize} 
\tablecaption{Photon Flux of Selected WR Stars in the MCs}
\tablehead{
\multicolumn{1}{l}{~~~~~~WR $\#$} & 
\multicolumn{1}{l}{WR Name} & 
\multicolumn{1}{c}{Observation Date} & 
\multicolumn{1}{c}{Aimpoint and Detector} & 
\multicolumn{1}{c}{$t_{\rm exp}$} & 
\multicolumn{1}{c}{Counts} & 
\multicolumn{1}{c}{Photon Flux} \\ 
\multicolumn{1}{c}{} &
\multicolumn{1}{c}{} &
\multicolumn{1}{c}{} &
\multicolumn{1}{c}{} &
\multicolumn{1}{c}{(ks)} &
\multicolumn{1}{c}{(cnts)} &
\multicolumn{1}{c}{(phot~cm$^{-2}$~s$^{-1}$)}   
}

\startdata

LMC-WR\,19              & Brey\,16  & 1999-12-04 UT12:28 & ACIS-I I2 & 30.4 &  40$\pm$7 & (2.0$\pm$0.4)$\times$10$^{-6}$~~ \\
                        &           & 2003-02-07 UT01:51 & ACIS-S S2 & 29.6 & 70$\pm$10 & (4.4$\pm$0.6)$\times$10$^{-6}$~~ \\
\hline
LMC-WR\,20              & Brey\,16a & 1999-12-04 UT12:28 & ACIS-I I2 & 30.4 &  51$\pm$8 & (2.3$\pm$0.4)$\times$10$^{-6}$~~ \\
                        &           & 2003-02-07 UT01:51 & ACIS-S S2 & 29.6 &  20$\pm$7 & (1.1$\pm$0.4)$\times$10$^{-6}$~~ \\
\hline
LMC-WR\,78              & Brey\,65b & 2000-12-07 UT19:26 & ACIS-S S2 & 91.7 &  11$\pm$3 & (2.1$\pm$0.9)$\times$10$^{-7}$~~ \\
                        &           & 2001-04-25 UT16:48 & ACIS-S S3 & 13.8 &   9$\pm$4 & (7.4$\pm$3.7)$\times$10$^{-7}$~~ \\
                        &           & 2002-05-15 UT05:55 & ACIS-S S3 & 44.2 &   9$\pm$5 & (3.0$\pm$1.8)$\times$10$^{-7}$~~ \\
\hline
LMC-WR\,85              & Brey\,67  & 2000-12-07 UT19:26 & ACIS-S S2 & 91.7 &100$\pm$12 & (1.92$\pm$0.23)$\times$10$^{-6}$~~ \\
                        &           & 2001-04-25 UT16:48 & ACIS-S S4 & 13.8 &  15$\pm$5 &   (1.9$\pm$0.6)$\times$10$^{-6}$~~ \\
                        &           & 2001-12-12 UT14:59 & ACIS-S S2 & 49.1 &  27$\pm$8 & (1.20$\pm$0.34)$\times$10$^{-6}$~~ \\
\enddata
\end{deluxetable}
}

\clearpage

\noindent
Figure 1  \\
(a) 
\emph{Chandra} ACIS X-ray images in the 0.5-7.0 keV band and optical 
images overlaid with X-ray contours for 
(top-left) Brey\,16 and Brey\,16a, 
(top-right) Brey\,56, 
(bottom-left) Brey\,65b, Brey\,57, and Brey\,65c, and  
(bottom-right) Brey\,66.  
Brey\,58 and TSWR\,4, close to a bright X-ray source, are also shown.  
The positions of the WR stars given by \citet{BAT99} and \citet{MOP03} 
are marked with a '+' sign.  
\emph{HST} broad-band images were used for Brey\,56 and Brey\,58 
(PI: Kirshner), and for Brey\,57, Brey\,65, Brey\,65b, and Brey\,65c 
(PI: Casertano), otherwise DSS optical images were used.  
The contour levels have been chosen to highlight the X-ray emission 
and identification of WR stars.   \\
(b) 
Same as Fig.~1a, for 
(top-left) Brey\,67, 
(top-right) Brey\,72, 
(bottom-left) Brey\,74a, and 
(bottom-right) R\,140a, R\,140b, and Brey\,86.  
\emph{HST} broad-band images were used for Brey\,72 (PI: Rhoads), 
Brey\,74a (PI: Mignani), and R\,140a, R\,140b, and Brey\,86 (PI: 
Trauger, Walborn, Westphal); 
DSS optical image was used for Brey\,67.  \\
(c)
Same as Fig.~1a, for 
(top-left) Brey\,89, 
(top-right) Brey\,90, 
(bottom-left) Brey\,94, Brey\,95, and Brey\,95a, and 
(bottom-right) HD\,5980.  
An \emph{HST} broad-band image was used for Brey\,90 (PI: Trauger, 
Walborn, Westphal), otherwise DSS optical images were used.  \\
(d)
Same as Fig.~1a, for Brey\,75, Brey\,76, Brey\,77, Brey\,78, Brey\,84, 
Mk\,35, and R\,136c.  
An \emph{HST} broad-band image was used for the optical image 
(PI: Trauger, Walborn, Westphal).  
R\,136b and Brey\,83, close to bright X-ray sources, are also shown.  \\
(e)
\emph{Chandra} ACIS-S (left) and ACIS-I (center) X-ray images in the 
0.5-7.0 keV band and DSS optical (right) images overlaid with 
ACIS-S X-ray contours of 
(top) Sk\,106 and 
(bottom) AV\,336a. \\  

\noindent
Figure 2 \\
\emph{Chandra} ACIS X-ray spectra of the WR stars in the MCs with 
sufficient number of counts over-plotted with the best-fit models 
listed in Tab.~6.  
For plotting purposes, the bin-width is 
300 eV for Brey\,16, Brey\,16a, Brey\,56, Brey\,78, and HD\,5980, 
150 eV for Brey\,67, R\,136a, and R\,136c, 
120 eV for R\,140a and Brey\,84, and 
 75 eV for Sk\,108 and AV\,336a.  
The lower panels show the residuals of the fit ($\Delta {\rm I}$) in terms 
of the bin sigmas ($\sigma$).  
The error bars in both the spectra and residual plots are 1-$\sigma$.  
\\

\noindent
Figure 3  \\
Combined \emph{Chandra} ACIS BI and FI spectra of the low-count 
WR stars in the MCs, with $\le$70 cnts, over-plotted with the 
best-fit model. For plotting purposes, the bin-width is 150 eV.  \\

\noindent
Figure 4 \\
X-ray luminosity versus \emph{Chandra} ACIS count rate for the WR 
stars in the MCs.
The filled circles correspond to the high-count subsample of WR 
stars in the MCs, while the open circles are WR stars in the low-count
subsample.  
The solid line corresponds to the best linear fit between X-ray 
luminosity and \emph{Chandra} ACIS count rate.  \\

\noindent
Figure 5  \\
Combined \emph{Chandra} ACIS BI and FI spectra of the WR stars in the 
MCs that are individually undetected over-plotted with the best-fit model. 
For plotting purposes, the bin-width is 300 eV for the ACIS BI spectrum 
and 150 eV for the ACIS FI spectrum.   \\

\noindent
Figure 6 \\
({\it top}) Count rate and ({\it bottom}) X-ray luminosity distributions 
of high-count (solid histogram) and low-count (dotted histogram) WR 
stars in the MCs.  
The vertical dashed line in the top panel marks the averaged count 
rate of the undetected sources described in \S3.2.3.  \\

\noindent
Figure 7 \\
Light curves of the WR stars in the MCs detected in X-rays with 
sufficient number of counts and time coverage.  
Count rates have been converted to photon fluxes to allow a fair comparison 
among observations of a WR star obtained with different detectors and 
offsets from the telescope aimpoint.  
Sk\,108 and AV\,336a have multiple observations, but while observations close 
in time of Sk\,108 had similar flux levels and were combined to improve the 
data quality, the multiple observations of AV\,336a show variations and were 
folded with its orbital period.   
The dotted sinusoidal curves over-plotted on the light curves of 
Brey\,84 and AV\,336a represent tentative fittings.

\end{document}